\documentclass[fleqn,10pt,twocolumn]{wlscirep}
\usepackage[utf8]{inputenc}
\usepackage[T1]{fontenc}

\usepackage{graphicx}%
\usepackage{multirow}%
\usepackage{amsmath,amssymb,amsfonts}%
\usepackage{amsthm}%
\usepackage{mathrsfs}%
\usepackage[title]{appendix}%
\usepackage{xcolor}%
\usepackage{textcomp}%
\usepackage{manyfoot}%
\usepackage{booktabs}%
\usepackage{algorithm}%
\usepackage{algorithmicx}%
\usepackage{algpseudocode}%
\usepackage{listings}%
\usepackage{siunitx}
\usepackage{longtable}
\usepackage[section]{placeins}
\usepackage[export]{adjustbox}
\usepackage{lineno}

\usepackage{todonotes} 
\let\oldtodo\todo
\renewcommand{\todo}[1]{\oldtodo[color=orange!40,inline]{#1}}

\title{RAAC panels can suddenly collapse before any warning of corrosion-induced surface cracking}

\author[1,2]{Ev\v{z}en Korec}

\author[3,4]{Peter Grassl}

\author[5]{Milan Jir\'{a}sek}

\author[1]{Hong S. Wong}

\author[2,*]{Emilio Mart\'{\i}nez-Pa\~neda}

\affil[1]{Department of Civil and Environmental Engineering, Imperial College London, Exhibition Road, London, SW7 2AZ, United Kingdom}

\affil[2]{Department of Engineering Science, University of Oxford, Parks Road, Oxford, OX1 3PJ, United Kingdom}

\affil[3]{James Watt School of Engineering, Rankine Building, Glasgow, G12 8LT, United Kingdom}

\affil[4]{Glasgow Computational Engineering Centre (GCEC), University of Glasgow, Glasgow, G12 8LT, United Kingdom}

\affil[5]{Department of Mechanics, Faculty of Civil Engineering, Czech Technical University in Prague, Th\'{a}kurova 7, Prague - 6, 166 29, Czech Republic}

\affil[*]{emilio.martinez-paneda@eng.ox.ac.uk}

\begin{abstract}
The collapse of reinforced autoclaved aerated concrete (RAAC) panels has attracted considerable public and academic interest. As detailed experimental data are not yet available and replicating the natural corrosion process requires years or decades, computational modelling is essential to understand under which conditions corrosion remains concealed. The very high porosity of RAAC is widely suspected to be a major contributing factor. However, current corrosion-induced cracking models are known to struggle with capturing the role of concrete porosity. To remedy this critical deficiency, we propose to enrich corrosion-induced cracking modelling with the analytical solution of reactive transport equations governing the precipitation of rust and a porosity-dependent description of diffusivity. With this, the corrosion concealment in RAAC panels is studied computationally for the first time, revealing that RAAC panels can suddenly collapse before any warning of corrosion-induced surface cracking and allowing to map the conditions most likely to result in sudden collapse.
\end{abstract}
\begin{document}

\flushbottom
\maketitle
%
%
\thispagestyle{empty}
\section*{Introduction}
\label{introRAAC}
The structural integrity of ageing RAAC structures has been a concern since multiple RAAC roof planks collapsed in the 1980s due to the corrosion of their embedded steel reinforcement \cite{Building_Research_Establishment1996-uh,Ball2023,currie1996reinforced}. These concerns stayed out of the public spotlight until 2018, when the Singlewell Primary School roof in Kent (UK) collapsed without warning. Although no one was injured as the collapse occurred during the weekend, over a hundred UK schools were ordered to shut just before the scheduled start of the 2023/2024 academic year. These events caused great inconvenience and public disconcert and there has been an unprecedented surge of public and academic interest in the corrosion-induced degradation of aerated concrete ever since. 

RAAC is an acronym for reinforced autoclaved aerated concrete and its origins can be traced back to architect Axel Eriksson in the 1920's in Sweden \cite{Ball2023}. Aerated concrete consists of cement, sand, sometimes lime and an aerating agent such as aluminium powder which generates hydrogen bubbles giving rise to a multitude of pores of typically millimetre diameter \cite{Thomas2023}. This mixture is often reinforced with steel rebars and steam-cured under high pressure (autoclaved). The resulting RAAC is highly porous, providing great advantages for the construction industry: it is lightweight, a quarter to a third of traditional concrete, and an excellent thermal and sound insulator. For these merits, aerated concrete became highly popular in the UK and Europe between the 1950s and 1990s \cite{Thomas2023}, especially for horizontal roof panels, but also pitched roofs, floors and wall panels \cite{Goodier2022}. Thus, today it can be found in many public buildings including schools, hospitals and courthouses \cite{provis2024material}. 

However, the high porosity which makes RAAC so attractive comes at a cost. Compared to traditional concrete, compressive, flexural, shear and tensile strength are considerably diminished, while creep is pronounced causing larger long-term deflections. Another downside stems from RAAC's high permeability \cite{Ball2023,Goodier2022}, which allows rapid ingress of water and other aggressive species resulting in several degradation mechanisms including freeze-thaw damage, and initiation of the corrosion of steel reinforcement \cite{Thomas2023,Liddell2023a}. For this reason, rebars are typically protected against corrosion by coating, but this protection also degrades in time. Although RAAC panel collapses are ultimately a multi-faceted issue which involves the role of overloading (changes in structural function), inadequate design (e.g. insufficient end bearing provision) and poor construction practices (e.g. insufficient anchorage from transverse steel, where panels are cut on site) \cite{Thomas2023,Liddell2023a}, corrosion-induced degradation has been arguably the greatest concern. 

A major challenge with RAAC is that the corrosion of steel reinforcement is difficult to detect. Intrusive techniques can easily damage low-strength aerated concrete and non-intrusive techniques such as penetrating radar are ineffective when applied through foil-backed insulation \cite{Liddell2023a}. In standard concrete, rebar corrosion typically manifests as corrosion-induced cracking and later spalling on concrete surfaces. Although corrosion-induced cracking is principally an adverse effect, it serves as an indicator and warning of ongoing rebar corrosion. However, the same does not hold for RAAC because its extensive pore space can accumulate a considerable amount of rust without exerting significant pressure on concrete, delaying corrosion-induced cracking significantly. This was noted in the report released by the Institution of Structural Engineers (IStructE), following the media upheaval, where it was stated that `...there are instances where intrusive surveys have shown corrosion of reinforcement has been advanced without any indication on the soffit of the panels'. If corrosion remains undetected and unchecked, it may result in a sudden collapse. For this reason, it remains critical to understand how long  corrosion can remain concealed or if it has to manifest before corrosion-induced collapse occurs. Although the microstructure and mechanical properties of aerated concrete have been subject of many studies \cite{Cabrillac2006, Narayanan2000, Qu2017, Narayanan2000a, Michelini2023, Trunk1999}, systematic on-site or experimental data on time-to-cracking of aerated concrete are still missing, stressing the need for reliable computational models. 

Even though there is currently no model for the time-to-cracking of aerated concrete, there have been a number of corrosion-induced time-to-cracking models proposed for standard concrete over the last decades, starting with the model of Bažant \cite{bazant1979physical} in 1979. Most of these are based on the idea of isolating a thick-walled concrete cylinder surrounding a single concrete rebar with the thickness of the cylinder equivalent to the thickness of the concrete cover of the rebar. As the review of Liang and Wang\cite{Liang2020} reports, the main differences between these models are their constitutive mechanical assumptions and the way they take into account emerging cracks. Some models \cite{ElMaaddawy2007,bazant1979physical,Liu1999a} assume only elastic behaviour with time-to-cracking being understood as the moment when average tensile stress equals concrete tensile strength. Within this approach, the stress in the tangential direction is sometimes considered to vanish after reaching the tensile strength \cite{Lu2011}.   
Other researchers \cite{Bhargava2005,Chernin2010,pantazopoulou2001modeling} took cracks into account by smearing them uniformly over the cracked zone. Finally, some studies explicitly considered the discrete crack morphology \cite{wang2004modelling,Balafas2011,Wang2004,Su2015}. Although all of these contributions are valuable, comprehensive reviews \cite{Chen2018,Reale2012,Jamali2013a} indicate that predictions of tested models are unreliable with the models of El Maaddawy and Soudki \cite{ElMaaddawy2007} and Chernin et al. \cite{Chernin2010} sometimes mentioned to provide the most reasonable results \cite{Liang2020,ElMaaddawy2007}. 

Reviews \cite{Chen2018,Reale2012,Jamali2013a} suggest that one of the most important reasons behind the lack of reliable models is that, regardless of the sophistication of the mechanical description of concrete cracking, predicted time-to-cracking is inherently strongly dependent on the porosity of concrete, which is typically taken into account with a porous zone around the steel rebar. Corrosion-induced pressure is commonly simulated by a rebar volumetric expansion and the porous zone is a free space in which the rebar can expand without inducing any pressure on the concrete. This simplified concept takes into consideration that the concrete pore space initially accommodates a substantial portion of precipitating rust \cite{Chitty2005,Care2008,Wong2010a,Taheri-Shakib2024} without inducing significant pressure on the concrete, delaying cracking. The problem with this concept is that the porous zone thickness, to which models are very sensitive, is an entirely fictitious parameter obtained by questionable estimates or fitting. This makes the application of current models to the time-to-cracking of aerated concrete impossible because there is no reliable way to extrapolate the thickness of a physically non-existent porous zone. 

As an alternative, Fahy et al. \cite{Fahy2017} proposed a coupled transport-structural approach, within which the transport of corrosion products from a steel-concrete interface is determined by a pressure gradient generated by the geometric confinement of rust. Furthermore, a new class of more complex corrosion-induced cracking models based on phase-field fracture   \cite{Fang2023,Pundir2023,Wei2021,Hu2022,Freddi2022,Korec2023,Korec2024,Korec2024a,Korec2024b} and other approaches such as neural network modelling \cite{Guneyisi2015} have been proposed. While these sophisticated models can handle arbitrary geometries and resolve the kinetic phenomena underpinning corrosion-induced cracking, the thick-walled-cylinder-based modelling approach provides an efficient and computationally robust avenue to obtain accurate estimates of time-to-cracking with minimum assumptions and input parameters. In the context of RAAC panel collapse, the enrichment of the thick-walled cylinder model with a suitable description of porosity and corrosion kinetics enables quantitative predictions that are currently not possible with experimental means over the time scales required.

To harness the numerical advantages of thick-walled cylinder models and overcome current limitations related to the concept of porous zone, we build upon our previous work \cite{Fahy2017,Aldellaa2022,Korec2023,Korec2024} and propose to replace the porous zone paradigm with a new physics-based approach based on the analytical solution of the reactive transport equations (with porosity-dependent diffusivity) governing the precipitation of rust in pore space and a dense rust layer. Within the proposed model, all parameters have a direct physical meaning and can be measured. With this new model, the corrosion concealment in RAAC panels is studied for the first time. 
The proposed model delivers predictions based on the concrete density, concrete cover thickness, and rebar diameter and thus paves the way for the future identification of the RAAC panels most endangered by sudden collapse without any warning of corrosion-induced surface cracking. In addition, new insight is gained into the scaling of critical corrosion penetration with concrete porosity.

\section*{Results}
\label{Results}
\subsection*{Impact of concrete porosity on the emergence of the first corrosion-induced surface crack}
\label{Sec:GenAspModRAAC}

\begin{figure*}[htp]
\begin{center}
\includegraphics[scale=0.58]{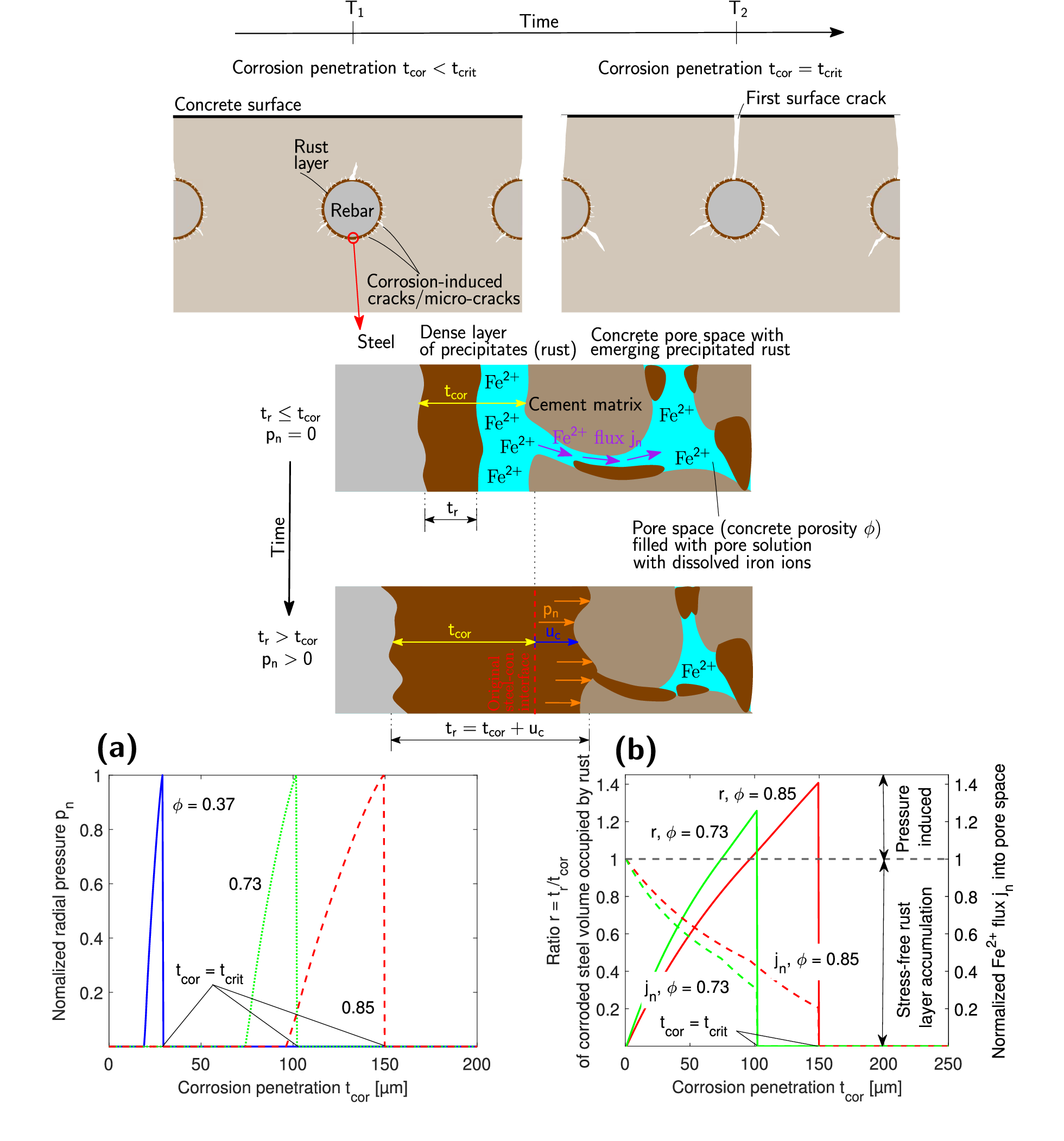}
\caption{ \textbf{The impact of concrete porosity on critical corrosion penetration.} (a) Critical corrosion penetration $t_{crit}$ for the first cracks on the concrete surface increases with concrete porosity $\phi$. This is because iron ions diffusivity increases with porosity and thus the normalised flux of iron ions into pore space $j_{n} $ decreases more slowly with corrosion penetration $t_{cor}$ (b), hindering the accumulation of a dense rust layer in the process. Pressure evolution comprises the stress-free period when accumulated rust fills the volume vacated by steel corrosion so that the thickness of the rust layer $t_{r}$ is smaller than the available thickness of the corroded layer $t_{cor}$. In the subsequent period, when $ t_{r}/t_{cor} > 1$, corrosion-induced pressure on concrete increases causing cracking in the vicinity of steel rebar until the dominant crack rapidly propagates to the concrete surface. In (a) and (b), concrete cover $c = 30$ mm and rebar diameter $d = 10$ mm were considered.}
\label{fig:genAspModP6}
\end{center} 
\end{figure*}

The time for the propagation of corrosion-induced cracks to the concrete surface is well-known to increase significantly with increasing concrete porosity. Assuming uniformly corroding steel rebar, this could be understood as the increase of the critical corrosion penetration $t_{crit}$ with concrete porosity $\phi$, where $t_{crit}$ is the thickness of the rust layer necessary to induce pressure sufficient for the propagation of corrosion-induced cracks to the concrete surface (see Fig. \ref{fig:genAspModP6}). This is because a significant portion of iron ions is transported to the concrete pore space, as observed in microscopy investigations \cite{Chitty2005,Care2008,Wong2010a}, reducing the number of ions contributing to the formation of a dense rust layer at the steel-concrete interface and thus also corrosion-induced pressure in the process. To study this behaviour, we developed a mathematical model discussed in the Methods section that allows to capture the impact of concrete porosity on $t_{crit}$, as can be observed in Fig. \ref{fig:genAspModP6}a. The discussion of the parameter values of the model and its calibration are provided in the Supplementary Information. The underlying reason for the increase of $t_{crit}$ in Fig. \ref{fig:genAspModP6}a is that the proposed model considers the well-known increase of diffusivity of ionic species in concrete with its porosity. The higher diffusivity in turn increases the flux $J_{II} $ of iron ions escaping into the concrete pore space (see Fig. \ref{fig:genAspModP6}b). Flux $J_{II} $ represents a portion of Faradays's law governed iron ions flux $ J_{II,Far} $ released from the steel surface. Thus, it is convenient to introduce the normalized flux $ j_{n} = J_{II}/J_{II,Far}$ indicating the portion of iron ions escaping into the pore space and thus not contributing to the buildup of the dense rust layer. See the Methods section for details on $J_{II}$ calculation. With the increasing thickness of the dense rust layer, $j_{n}$ declines steadily and an increasingly larger portion of iron flux accumulates in the dense rust layer. The higher the concrete porosity, the more gradual the decrease of $j_{n}$. In Fig. \ref{fig:genAspModP6}b, it can also be observed that $t_{crit}$ is the sum of two contributions. The first one is related to the stage during which the volume of accumulated dense rust layer can be accommodated stress-free by the volume of corroded steel, so that the hypothetical rust expansion ratio $r = t_{r}/t_{cor} \leq 1 $, where $t_{r} $ is the thickness of the rust layer and $t_{cor}$ is the thickness of the corroded steel layer (see Fig. \ref{fig:genAspModP6}). Upon reaching $ r = 1 $, the dense rust layer starts to exert pressure on concrete, which induces cracking understood as smeared axially uniform damage in the vicinity of the rebar. This lasts until the eventual propagation of a main crack through the concrete cover, at which point cracking can no longer be reasonably represented as uniformly smeared in the vicinity of the steel-concrete interface, and the proposed model thus cannot be applied anymore. The magnitude of contributions to $t_{crit}$ in both stages of pressure evolution increases with concrete porosity.

\subsection*{Critical corrosion penetration for aerated concrete}
\label{Sec:RAAC}

\begin{figure*}[htp]
\begin{center}
\includegraphics[scale=0.55]{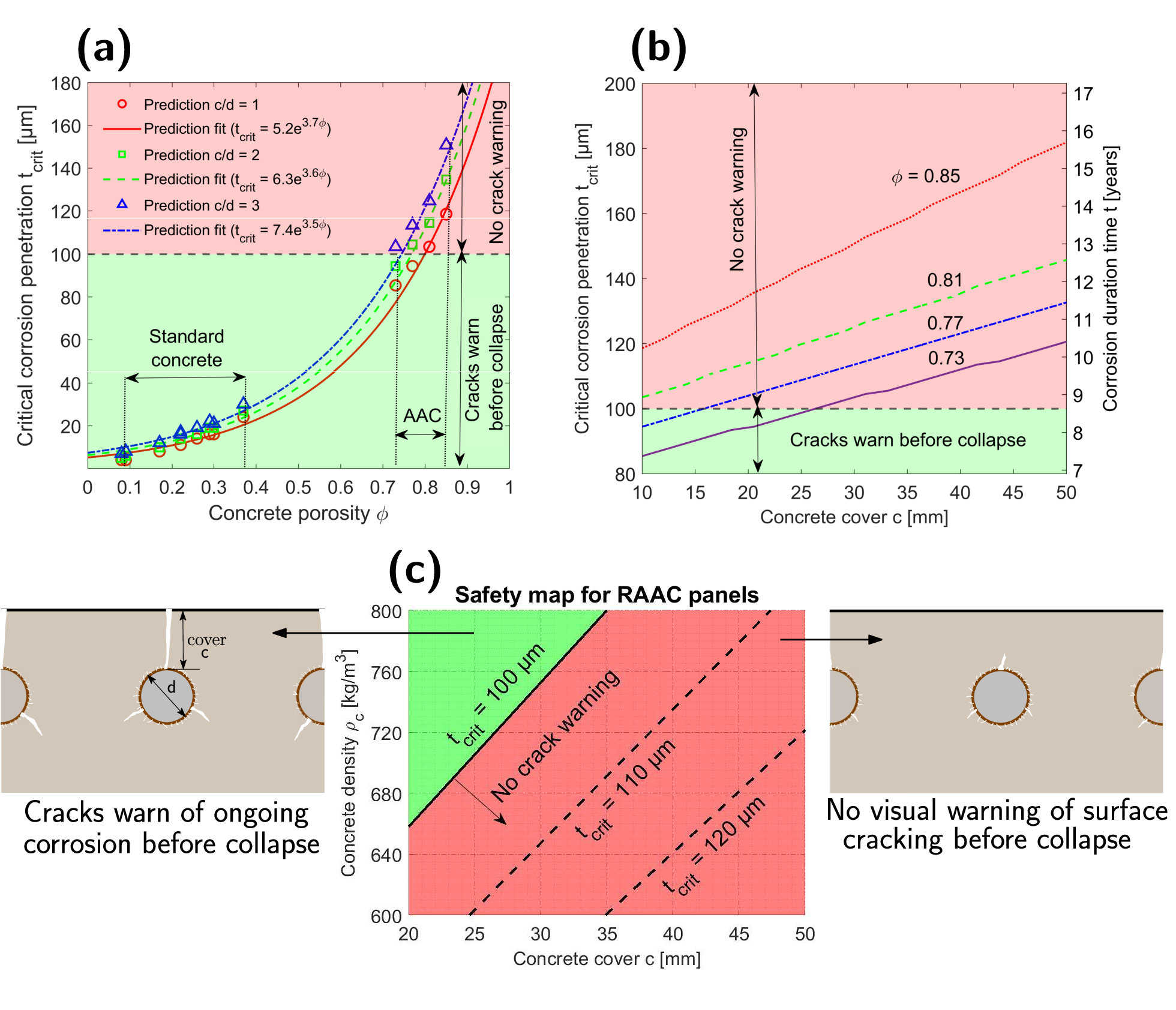}
\caption{\textbf{The critical corrosion penetration for RAAC panels.} (a) The critical corrosion penetration $t_{crit}$ increases exponentially with concrete porosity. A larger $c/d$ ratio (concrete cover to steel rebar diameter) and the thickness of a concrete cover also increases $t_{crit}$ (b). Excessively large $t_{crit}$ of aerated concrete means that steel corrosion in RAAC panels can be concealed well over 8 years, considering a typical value of corrosion current density about 1 \unit{\micro\ampere\per\centi\metre^2} \cite{Otieno2012a, Otieno2016a, Andrade2023, Walsh2016, andrade2023role} (b). We can see that $t_{crit}$ in aerated concrete can easily exceed 100 microns at which rebar-concrete bond in standard concrete was found to deteriorate to the extent that the ultimate limit state may be affected \cite{Andrade1993}. Although the exact threshold value of $t_{crit}$ for RAAC is not known yet, this indicates that RAAC panels can potentially collapse before any visually detectable warning of corrosion-induced surface cracking. The model allows to assess whether the panel with a given concrete density (which is directly linked to its porosity) and the thickness of a concrete cover is likely to be at the greatest risk of collapse before surface cracking (c) and thus narrow down and prioritise inspections or repairs. The safety map in (c) was calculated assuming steel rebar diameter $d = 10$ mm.}
\label{fig:RAACP6}
\end{center} 
\end{figure*}

The analysis of predicted critical corrosion penetrations $t_{crit}$ for samples from both standard and aerated concrete (with properties discussed in the Supplementary Information, particularly Supplementary Table 3) revealed that $t_{crit}$ increases exponentially with concrete porosity $\phi$ (see Fig. \ref{fig:RAACP6}a). As can be expected, $t_{crit}$ increases with the thickness of the concrete cover (see Figs. \ref{fig:RAACP6}b and \ref{fig:RAACP6}c) and the larger the porosity, the larger the rate of increase. 

While $t_{crit}$ in standard concrete is typically about few tens of microns (see results of experimental studies from Refs.\cite{AldellaaThesis2024,Vu2005,Al-Harthy2011,Lu2011} summarized in the Supplementary Table 3 or Ref. \cite{Andrade1993}), numerical predictions suggest that $t_{crit}$ can easily increase well over 100 microns for aerated concrete. If we consider the typical value of corrosion current density in naturally corroding standard reinforced concrete to be about 1 \unit{\micro\ampere\per\centi\metre^2} \cite{Otieno2012a, Otieno2016a, Andrade2023, Walsh2016, andrade2023role}, this means that steel corrosion in RAAC panels can be concealed well over 8 years (see Fig. \ref{fig:RAACP6}b), which agrees with on-site observations \cite{Liddell2023a}. 

Surface cracking is a very important indicator of otherwise difficult to detect steel corrosion. For this reason, it is a concern in RAAC whether corrosion can proceed undetected for so long that it leads to structural collapse prior to corrosion-induced cracks appearing on the concrete surface. With ongoing rebar corrosion, an increasingly thicker dense rust layer acts as a weak interface between the rebar and concrete. This, together with corrosion-induced cracking at the steel-concrete interface, negatively affects the bond between steel and concrete \cite{Feng2016a,Desnerck2015}. If unchecked, this process ultimately results in corrosion-induced structural collapse. Andrade et al.\cite{Andrade1993} reported that a bar cross-section loss of up to 100 microns is not expected to seriously affect the ultimate limit state of elements from standard reinforced concrete. Even though no similar estimates for aerated concrete are available, in Figs. \ref{fig:RAACP6}a, \ref{fig:RAACP6}b and \ref{fig:RAACP6}c we can see that the threshold of 100 microns was overcome in many of the performed case studies. This suggests that there is a risk of RAAC panels collapsing as a result of corrosion-induced degradation before any cracking on the concrete surface occurs. The proposed model allows the calculation of safety assessment maps (see Fig. \ref{fig:RAACP6}c) indicating the most endangered cases based on the panels' density (directly linked to their porosity), the thickness of a concrete cover and the diameter of the steel bar. This illustrates how computational modelling can help to determine the most vulnerable panels so that their inspections or repairs can be prioritised. As there is a notorious lack of measurements of 
transport and chemical parameters for RAAC, potential variations in the predictions of $t_{crit}$ due to the uncertainties associated with parameter values, especially reaction rates, are discussed in the Supplementary Information (see Supplementary Fig. 1). Let us also note that the axisymmetric nature of the proposed model allows us to investigate only the case of uniform corrosion. However, it is well-known that rebar corrosion can be non-uniform on the steel surface, especially in the case of chloride-induced corrosion. In this case, crack evolution would be delayed \cite{Korec2024}, enlarging $t_{crit}$ in the process. As many RAAC panels can be expected to be strongly carbonated \cite{Liddell2023a} and not exposed to high chloride concentrations, the adopted assumption of uniform corrosion is considered realistic. However, except for the report by Liddell et al. \cite{Liddell2023a}, there are currently no available studies allowing an assessment of the non-uniformity of corrosion in RAAC panels exposed to real conditions and thus it has to be remembered that $t_{crit}$ can be even larger than predicted by the presented model, leading to an even longer and thus more dangerous corrosion concealment.

\section*{Discussion}\label{secDiscussion}

In this study, a new chemo-mechanical model for corrosion-induced cracking of reinforced concrete was proposed. By taking into account the porosity-dependent diffusivity of concrete and the reactive transport equations governing the precipitation of rust, it allows to consider the critical impact of concrete porosity on the period over which corrosion does not induce any cracks on the concrete surface and thus remains dangerously concealed. The model was calibrated with the available data on critical corrosion penetration from tests conducted on standard concrete samples (as data for RAAC are not yet available). Furthermore, for the first time, the critical corrosion penetration of reinforced autoclaved aerated concrete (RAAC) relevant to recent public concern over the safety of RAAC panels was studied. The proposed model allows to predict the critical corrosion penetration of RAAC panels based on their density, the thickness of the concrete cover and the rebar diameter. Thus, it demonstrates that computational modelling can help engineers dealing with RAAC to identify panels at the greatest risk of collapse before any warning of surface cracking and thus better target inspections and repairs. The main findings can be characterised as follows: 
\begin{itemize}
\item Critical corrosion penetration $t_{crit}$ was found to increase exponentially with concrete porosity $\phi$ (see Fig. \ref{fig:RAACP6}a). Thus, $t_{crit}$ was found to be well over 100 microns for many simulated reinforced autoclaved aerated concrete (RAAC) specimens. This is equivalent to over 8 years of ongoing concealed corrosion (i.e. without any manifestation of surface cracking) if corrosion current density of 1 \unit{\micro\ampere\per\centi\metre^2} typical for reinforced concrete in natural conditions \cite{Otieno2012a, Otieno2016a, Andrade2023, Walsh2016, andrade2023role} is assumed (see Fig. \ref{fig:RAACP6}b). 
\item During the corrosion of steel rebars, the bond between steel and concrete is damaged because of the presence of a weak interface layer of brittle corrosion products accumulated in a dense rust layer at the steel-concrete interface and local concrete cracking. As the predicted $t_{crit}$ for aerated concrete was found to be well over 100 microns in many investigated case studies, literature on standard concrete \cite{Andrade1993} suggests that at such a corrosion penetration, the steel-concrete bond is deteriorated to such an extent that an ultimate limit state is significantly affected. This indicates that corrosion-induced damage can cause RAAC panels to collapse before any visual indication of cracking on the concrete surface.     
\end{itemize}

\section*{Methods}\label{secMethods}

\subsection*{Calculation of the thickness of the dense rust layer and its pressure}
\label{sec:denLayCalc}

In the proposed model, corrosion-induced pressure is evaluated as the pressure of a dense rust layer evolving under geometrically constrained conditions in the space vacated by steel corrosion. The pressure resulting from the constrained accumulation of rust in pore space analysed in previous studies of the authors (see Refs. \cite{Korec2023,Korec2024}) is neglected. This simplifying assumption is considered reasonable for highly porous aerated concrete, which is the main interest of this study. Thus, to predict a pressure induced by a dense rust layer, the portion of rust accumulating in a dense rust layer (i.e. the portion of rust that did not escape into the concrete pore space) has to be evaluated. To this end, we employ a reactive transport model proposed in the previous work of the authors of this study \cite{Korec2024a}. 

After corrosion initiation, $ \mathrm{Fe}^{2+} $ ions are released from the steel surface into the pore solution and the amount of released ions is proportional to the corrosion current of density $i_a$ according to Faraday's law $J_{II,Far} = i_{a}/(z_{a}F)$, where $F$ is Faraday's constant and $ z_{a} = 2 $ is the number of electrons exchanged in anodic corrosion reaction. Then, ferrous ions undergo a complex series of chemical reactions (see Refs.\cite{Furcas2022,Wieland2023}), forming a number of intermediate products before eventually precipitating into rust. X-ray diffraction (XRD) measurements reveal that rust is composed mostly of iron oxides and iron hydroxy-oxides \cite{Chitty2005, Zhao2011b, Zhang2019c}. Thus, inspired by previous studies \cite{Stefanoni2018b,Zhang2021}, the reactions of the simplified chemical system are proposed to be the oxidation of $ \mathrm{Fe}^{2+} $ ions to $ \mathrm{Fe}^{3+} $ ions and their subsequent precipitation into iron oxides and iron hydroxy-oxides respectively. 

The portion of the flux of $ \mathrm{Fe}^{2+} $ ions escaping into pore space and thus not participating in the formation of the dense rust layer can be expressed as  
\begin{equation}
J_{II} = k_{f} J_{II,Far} = k_{f}\frac{i_{a}}{z_{a}F}
\end{equation}
where $k_{f}$ is the flux reduction coefficient. To estimate $k_{f}$ we follow Stefanoni et al.\cite{Stefanoni2018b} and assume diffusion-dominated transport of iron ions and an instantaneous precipitation of $ \mathrm{Fe}^{3+} $ ions. Importantly, there are two domains -- rust and concrete (see Fig. \ref{fig:genAspModP6}), with the rust domain changing its size in time. Both domains are characterised by different diffusivity of $ \mathrm{Fe}^{2+} $ ions in rust $ D_{r} $ and concrete $D_{c}$ which is assumed to be constant in time. The diffusivity of iron ions in concrete $D_{c}$ strongly depends on the concrete porosity. Thus, in this model, we assume that $D_{c} = D_{w}\phi^{m}$ where $D_{w}$ is the diffusivity of $ \mathrm{Fe}^{2+} $ ions in water and $\phi$ is the total porosity of concrete. Exponent $m$ was calibrated based on experimental data from studies\cite{AldellaaThesis2024,Vu2005,Al-Harthy2011,Lu2011} on standard concrete as discussed further (see Supplementary Fig. 2). Let us note here that $D_{c}$ is in reality also affected by cracking at the steel-concrete interface and the clogging of pores by rust. In the proposed model, the averaged impact of this influence can be understood to be encapsulated in exponent $m$. 

The time-dependent thickness of the rust domain is $t_{cor} + u_{c}$, where $t_{cor}$ is the thickness of the rust layer and $u_{c}$ is the unknown displacement of the steel-concrete boundary due to pressure induced by the accumulating dense rust layer. The thickness of concrete domain $t_{c}$ represents a maximum distance from the steel-concrete boundary to which $ \mathrm{Fe}^{2+} $ can be transported in concrete pore space. When simplifying the solution domain to be one-dimensional, the reactive transport problem for the concentration of $ \mathrm{Fe}^{2+} $ ions $c_{II}$ on the domains of rust and concrete thus reads
\begin{subequations}\label{statDiffEq}
\begin{align}
\label{statDiffEqRust}
&D_{r}\dfrac{\mathrm{d}^{2} c_{II}}{\mathrm{d} x^{2}} = k^{II \rightarrow o}_{r} c_{II} + k^{II \rightarrow III}_{r} c_{II}c_{ox}  \\ 
&\text{in } \,\, x \in \left\langle 0,t_{cor}+u_{c} \right) \nonumber \\
\label{statDiffEqCon}
&D_{c}\dfrac{\mathrm{d}^{2} c_{II}}{\mathrm{d} x^{2}} = k^{II \rightarrow o}_{r} c_{II} + k^{II \rightarrow III}_{r} c_{II}c_{ox} \\
&\text{in } \,\, x \in \left\langle t_{cor}+u_{c}, t_{cor}+u_{c}+t_{c} \right\rangle \nonumber
\end{align}
\end{subequations}
where term $k^{II \rightarrow o}_{r} c_{II}$ and $k^{II \rightarrow III}_{r} c_{II}c_{ox}$ represent the precipitation of $ \mathrm{Fe}^{2+} $ ions and the transformation of $ \mathrm{Fe}^{2+} $ to $ \mathrm{Fe}^{3+} $ ions respectively. Parameters $ k^{II \rightarrow III}_{r} $ and  $k^{II \rightarrow o}_{r}$ are the corresponding reaction constants and $c_{ox}$ is an oxygen concentration. Assuming the continuity of flux between steel and concrete domain, Faraday's law dictated flux $J_{II,Far}$ at $x = 0$, $ c_{II} = 0 $ at $x = t_{cor}+u_{c} $, equations (\ref{statDiffEqRust}) and (\ref{statDiffEqCon}) can be solved analytically and the flux reduction coefficient $k_{f}$, which reads 
\begin{equation}
k_{f} = \dfrac{2{\rm e}^{
A_{r}} \sqrt{D_{c}} \coth A_c}{\left(1+{\rm e}^{2A_{r}}\right)\left(\sqrt{D_{c}} \coth A_c + \sqrt{D_{r}} \tanh A_r\right)}
\end{equation}
is calculated using rust-related and concrete-related constants 
\begin{equation}
A_{r} = t_{cor}\sqrt{\frac{k^{II \rightarrow o}_{r}+c_{ox} k^{II \rightarrow III}_{r}}{D_{r}}}, \quad A_{c} = t_{c}\sqrt{\frac{k^{II \rightarrow o}_{r}+c_{ox} k^{II \rightarrow III}_{r}}{D_{c}}}
\end{equation}
Iron ions that do not escape into pore space form a dense rust layer. The hypothetical thickness $t_{unc}$ of the unconstrained rust layer in time $T$ is  
\begin{equation}
t_{unc} = \kappa \left(t_{cor} - \frac{M_{Fe}}{\rho_{Fe}} \displaystyle \int_{0}^{T}k_{f} \frac{i_{a}}{z_{a}F} \mathrm{d} t \right)
\end{equation}
where $\kappa$ is the molar volume ratio of rust and steel, often denoted as the expansion coefficient of rust. Rust has a significantly lower density than steel. Thus, while rust that does not escape into pore space can initially accumulate freely (no mechanical pressure induced on concrete), at a certain point in time the dense rust layer becomes spatially constrained by concrete, which is the underlying reason of the rise of corrosion-induced pressure $ p $ leading to the radial displacement $ u_{c} $ of the original steel-concrete boundary (see Fig. \ref{fig:genAspModP6}). 


Pressure $ p $ can be calculated from thermodynamic considerations reflecting the finite compressibility of rust. Assuming constant bulk modulus of rust $K_r$ and isothermal conditions, $K_r$ can be expressed as  
\begin{equation}\label{compressibility}
\dfrac{1}{K_r} = -\dfrac{1}{V_{r,d}}\left(\dfrac{\partial V_{r,d}}{\partial p}\right)
\end{equation}  
where $ V_{r,d} = \pi((r_{i}+u_{c})^2-(r_{i}-t_{cor})^2)L $ is a volume occupied by rust layer. In these relations, $L$ is the length of the rebar and $r_{i}$ is the uncorroded rebar radius. In (\ref{compressibility}), the bulk modulus of rust is calculated as $  K_{r} = E_r/(3(1-2\nu_r))$ , with $E_r$ and $\nu_r$ being respectively the Young modulus and Poisson ratio of rust. By integration of (\ref{compressibility}), we arrive at  
\begin{equation}\label{compressibility5}
p = K_{r} \ln \dfrac{V_0}{V_{r,d}} 
\end{equation}    
where integration constant $ V_{0} = \pi((r_{i}+t_{unc})^2-(r_{i}-t_{cor})^2)L$ is the volume that would be occupied by the rust layer if it were not spatially constrained by concrete (i.e.\ under pressure $p=0$). The primary unknown variables of our problem are thus $p$ or $u_{c}$. If one of these is known, the other can be calculated. In addition to thermodynamic requirement (\ref{compressibility5}), pressure $ p $ also has to fulfill the conditions of mechanical equilibrium in the concrete domain. This second requirement is discussed next.

\subsection*{Cracked thick-walled cylinder model}
\label{sec:crkThcCylMod}

\begin{figure}[htp]
\begin{center}
\includegraphics[scale=0.85]{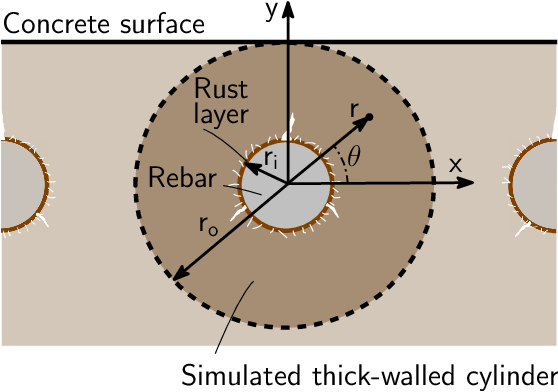}
\caption{\textbf{Graphical illustration of the reinforced concrete specimen with the geometry of a thick-walled concrete cylinder solution domain and polar coordinates.}}
\label{fig:crkWalldCyl}
\end{center} 
\end{figure}

Corrosion-induced pressure $ p $ of rust on concrete (whose origins are discussed in the previous section) has to be in equilibrium with the mechanical stress field in gradually cracking concrete. For the mechanical description of the concrete domain, the cracked thick-walled cylinder model of Grassl et al. is adopted \cite{Grassl2019,Fahy2017,Aldellaa2022}. The main underlying idea of this model is that a thick-walled cylinder surrounding steel rebar is virtually isolated from the concrete sample (see Fig. \ref{fig:crkWalldCyl}). The inner radius of the thick-walled cylinder $r_{i}$ coincides with the radius of uncorroded steel rebar and the outer radius $r_{o}$ represents the distance from the centre of steel rebar to the concrete surface. The radial pressure of rust layer $ p $ on the inner concrete boundary results in its radial concrete displacement $u_{c}$ and corrosion-induced damage. This firstly manifests as cracking or microcracking in the vicinity of a steel-concrete interface (see e.g. Ref. \cite{Taheri-Shakib2024}),  later followed by the rapid formation of cracks between the rebar and the concrete surface (see Figs. \ref{fig:genAspModP6} and \ref{fig:crkWalldCyl}). Assuming uniform corrosion, initial damage in the vicinity of the steel-concrete interface is thus smeared and also considered uniform before the propagation of cracks to a concrete surface. The assumption of uniform corrosion is considered to be realistic for RAAC panels not exposed to chlorides, as they are expected to be highly carbonated\cite{Liddell2023a} such that corrosion is initiated on the entire steel surface. Let us also assume that concrete is mechanically isotropic and linear elastic before cracking and neglect the effects of other loads and interaction with corrosion-induced cracks of surrounding rebars or free concrete surfaces (in the case of a corner-located rebar). With these assumptions, the problem of corrosion-induced fracture can be simplified to be axisymmetric and parametrized with radial distance $r$ (see Fig. \ref{fig:crkWalldCyl}). 

As the sample geometry, corrosion-induced load and material properties are uniform along the length of the rebar, our problem can be reduced to two-dimensional analysis assuming either plane strain or plane stress conditions. It is difficult to determine which one of these conditions is more accurate. While Chernin et al. \cite{Chernin2010} employed plane strain conditions expected in the bulk of reinforced concrete elements, Pantazopoulou and Papoulia\cite{pantazopoulou2001modeling} argued in favour of plane stress proposing that these elements are already cracked from other loads. Here, in accordance with previous studies \cite{Grassl2019, Aldellaa2022}, we employ plane stress conditions expecting that this choice is not likely to affect results significantly. 

Radial stress $ \sigma_{r} $ and tangential stress $\sigma_{\theta}$ have to fulfil equilibrium condition         
\begin{equation}\label{equEq}
\frac{\mathrm{d} \sigma_{\mathrm{r}}}{\mathrm{d} r} \,r+\sigma_{r}-\sigma_\theta=0
\end{equation}
The corresponding strain components $ \varepsilon_{r} $ and $ \varepsilon_{\theta} $ are calculated from radial displacement $u$ as   
\begin{equation}\label{kinEq}
\varepsilon_{r}=\frac{\mathrm{d} u}{\mathrm{~d} r} \quad \text { and } \quad \varepsilon_\theta=\frac{u}{r}
\end{equation}
and they are linked to stress components $ \sigma_{r} $ and $\sigma_{\theta}$ via extended Hooke's law 
\begin{equation}\label{HookLawExt}
\left\{\begin{array}{l}
\varepsilon_{\mathrm{r}} \\
\varepsilon_\theta
\end{array}\right\}=\frac{1}{E}\left(\begin{array}{cc}
1 & -\nu \\
-\nu & 1
\end{array}\right)\left\{\begin{array}{l}
\sigma_{\mathrm{r}} \\
\sigma_\theta
\end{array}\right\}+\left\{\begin{array}{c}
0 \\
\varepsilon_\theta^{\mathrm{cr}}
\end{array}\right\}
\end{equation}
where $E$ and $\nu$ are Young's modulus and Poisson's ratio of concrete, respectively, and $\varepsilon_\theta^{cr}$ is the cracking strain, i.e. the inelastic fracture-induced component of tangential strain. By expressing stresses ($\sigma_{r},\sigma_{\theta}$) in terms of strains ($\varepsilon_r,\varepsilon_\theta$) from (\ref{HookLawExt}) and substituting these and kinematic relations (\ref{kinEq}) into (\ref{equEq}), we obtain
\begin{equation}\label{govEq1}
\frac{\mathrm{d}^2 u}{\mathrm{~d} r^2}+\frac{1}{r} \frac{\mathrm{d} u}{\mathrm{~d} r}-\frac{1}{r^2} u+\frac{1}{r}(1-\nu) \varepsilon_\theta^{\mathrm{cr}}-\nu \frac{\mathrm{d} \varepsilon_\theta^{\mathrm{cr}}}{\mathrm{d} r}=0
\end{equation}
For $\sigma_{\theta} \leq f_{t}$, i.e. before the tangential stress reaches the value of tensile strength, it is considered that $\varepsilon_\theta^{cr} = 0$. For $ \sigma_{\theta} > f_{t} $, $\varepsilon_\theta^{cr} $ is calculated following the formula presented by Aldellaa et al.\cite{Aldellaa2022} who assumed an exponential softening law for the dependence of $\sigma_\theta$ on $\varepsilon_\theta^{\mathrm{cr}}$. For the sake of simplification and numerical robustness of the model, we will consider fracture energy $ G_{f} \rightarrow \infty $, which is equivalent to ideal plastic material behaviour upon reaching $\sigma_\theta = f_{t}$. Applying this assumption to the formula of Aldellaa et al.\cite{Aldellaa2022} yields
\begin{equation}\label{crStrEq}
\frac{\mathrm{d} \varepsilon_\theta^{\mathrm{cr}}}{\mathrm{d} r}=\frac{\mathrm{d} u}{\mathrm{~d} r} \frac{1}{r}-\frac{u}{r^2}+\nu \frac{\mathrm{d}^2 u}{\mathrm{~d} r^2}
\end{equation}
Substituting (\ref{crStrEq}) back to (\ref{govEq1}) leads to the ordinary differential equation 
\begin{equation}\label{govEq2}
\frac{\mathrm{d}^2 u}{\mathrm{~d} r^2}+\frac{1}{r(1+\nu)}\left( \frac{\mathrm{d} u}{\mathrm{~d} r} -\frac{1}{r} u + \varepsilon_\theta^{\mathrm{cr}}\right)=0
\end{equation}  
which together with (\ref{crStrEq}) constitutes the cracked thick-walled cylinder model. Boundary conditions for (\ref{govEq2}) are $u(r_{i})=u_{c}$ and $ \sigma_{r}(r_{o}) = 0$. The corrosion-induced displacement $u_{c}$ is unknown but it has to result in such a corrosion-induced pressure $ p = -\sigma_{r}(r_{i}) $ on the inner concrete boundary that is equal with pressure $p$ from formula (\ref{compressibility5}) reflecting rust compressibility. 

In this model, critical corrosion penetration $t_{crit}$, i.e. the corroded steel thickness leading to the first surface cracks, is considered to be the thickness of the corroded steel layer at which the circumferential stress $\sigma_{\theta}$ on the inner concrete boundary (i.e. at $r = r_{i}$) reaches its maximum value $\sigma_{\theta}(r_{i}) = f_{t}$. Upon reaching this threshold, the crack is assumed to rapidly propagate through the concrete cover (see Fig. \ref{fig:genAspModP6}). Although this condition is simplified, it was found to provide very reasonable estimates in comparison with detailed lattice modelling \cite{Aldellaa2022}. Equation (\ref{equEq}) implies that the criterion $\sigma_{\theta}(r_{i}) = f_{t}$ is equivalent with $p = f_{t}(r_{o}-r_{i})/r_{i}$, which allows to define a non-dimensional radial pressure $p_{n} = p(r_{i}/(f_{t}(r_{o}-r_{i})))$ so that $ p_{n} = 1 $ when $\sigma_{\theta}(r_{i}) = f_{t}$. For more details on the numerical solution and the discussion of the choice of the values of parameters, the reader is encouraged to see the Supplementary Information.  

\section*{Data availability}

The dataset used to calibrate the model is presented in the Supplementary Information (see Supplementary Table 3).

\section*{Code availability}

All developed codes are freely available at \url{https://mechmat.web.ox.ac.uk/}.


\section*{Acknowledgements}

E. Korec acknowledges financial support from the Imperial College President’s PhD Scholarships. M. Jirásek acknowledges financial support from the European Union under the ROBOPROX project (reg.~no.~CZ.02.01.01/00/22\_008/0004590).
E. Mart\'{\i}nez-Pa\~neda was supported by an UKRI Future Leaders Fellowship [grant MR/V024124/1]. We additionally acknowledge computational resources and support provided by the Imperial College Research Computing Service. 


\section*{Author contributions statement}

Conceptualisation: E.K., P.G., M.J., H.S.W., E.M.P. Methodology: E.K., P.G. Investigation: E.K., P.G., M.J., H.S.W., E.M.P. Visualisation: E.K.
Funding acquisition: E.K., M.J., H.S.W., E.M.P. Project administration: E.K., P.G., M.J., H.S.W., E.M.P. Supervision:
P.G., M.J., H.S.W., E.M.P. Writing—original draft: E.K. Writing—review and editing:
E.K., P.G., M.J., H.S.W., E.M.P.

\section*{Competing interests}
The authors declare no competing interests.

\section*{Additional information}
\textbf{Correspondence} and requests for materials should be addressed to Emilio Mart\'{\i}nez-Pa\~neda.

\end{document}